\documentclass[aps,prl,twocolumn,showpacs,superscriptaddress,groupedaddress]{revtex4}
\usepackage{bbm,amsmath,amssymb,amsbsy,graphicx,subfigure,times}

\usepackage{hyperref}
\usepackage[latin1]{inputenc}
\newcommand{\gr}[1]{\boldsymbol{#1}}
\newcommand{\ket}[1]{|#1\rangle}
\newcommand{\bra}[1]{\langle#1|}

\newcommand{\eq}[1]{Eq.~(\ref{#1})}

\newcommand{\sig}{{\gr\sigma}}

\begin{document}
\title{Quantum Benchmark for Teleportation and Storage of Squeezed States}

\date{April 2, 2008}
\author{Gerardo Adesso}
\affiliation{Dipartimento di Matematica e Informatica, Universit\'a degli Studi di Salerno, Via Ponte Don Melillo, 84084 Fisciano (SA), Italy;} \affiliation{Grup de F\'isica
Te\`orica, Universitat Aut\`onoma de Barcelona, 08193 Bellaterra
(Barcelona), Spain.}

\author{Giulio Chiribella}
\affiliation{QUIT Quantum Information Theory Group  and Dipartimento
di Fisica ``A. Volta'', University of Pavia, via A. Bassi 6, I-27100
Pavia, Italy}


\begin{abstract}
  We provide a quantum benchmark for teleportation and storage of
  single-mode squeezed states with zero displacement and completely
  unknown degree of squeezing along a given direction.
  For pure squeezed input states, a fidelity higher than $81.5\%$ has
  to be attained in order to outperform any classical strategy based
  on estimation of the unknown squeezing and repreparation of squeezed
  states.  For squeezed thermal input states, we derive an upper and a
  lower bound on the classical average fidelity, which tighten for moderate degree of mixedness.
  These results enable a critical discussion of recent experiments with squeezed light.
\end{abstract} \maketitle


Quantum teleportation \cite{bennett} and quantum state storage
\cite{storage} are among the pillars of quantum information science,
not only as groundbreaking demonstrations of the usefulness of
quantum correlations against limited classical scenarios, but also
as essential ingredients for quantum computation \cite{qc} and
long-distance quantum communication
\cite{repeaters}.
Continuous variable (CV) systems \cite{brareview}, where quantum
correlations arise between canonically conjugate observables (like
the quadratures of light or the collective spin components of atomic
ensembles), are especially suited for the
unconditional implementation of these protocols \cite{fururep}. While
 the seminal CV experiments only dealt with light
 fields \cite{furuscience},
progresses towards the establishment of a
 complete quantum interface
between light and matter have been
 recently undertaken, with
milestones like  the storage
 of
 coherent light onto atomic
memories \cite{memorypolzik}, and the
 teleportation from light to
matter \cite{telepolzik}.

 Quantum entanglement is the key
resource that allows a sender
 (Alice) and a receiver (Bob) to beat
any classical strategy for
 transmitting and storing quantum states.
In realistic implementations, however,
 the quality of
transfer/storage is limited by factors such as
 imperfections and
losses. To decide whether an experiment
 demonstrates a genuinely
quantum feature, one needs appropriate
 figures of merit and
appropriate benchmarks in terms of them. The
 typical figure of merit
characterizing quantum teleportation (from
 now on we restrict to this
protocol, bearing in mind that our
 results equally apply to state
storage too) is the {\em fidelity}
 \cite{Fidelity} ${\cal F} =
\{{\rm
  Tr}[(\sqrt{\varrho^{in}}\varrho^{out}\sqrt{\varrho^{in}})^{1/2}]\}^2$
between the unknown input state $\varrho^{in}$ to be teleported by
Alice and the
 output state $\varrho^{out}$ which is actually obtained
by Bob.
To conclude that a quantum demonstration has taken place, the
fidelity ${\cal F}$ has
 to beat the best possible fidelity ${\cal
  F}^{cl}$---usually called {\em classical fidelity threshold} (CFT)---achievable
 by two cheating parties who have access to
unlimited ``classical''
 means (local operations and classical
communication) but are not able to share entanglement, nor to
directly transmit quantum systems \cite{bfkjmo}. Under this
restriction, the only possibility for Alice and Bob is a
 ``measure-and-prepare'' strategy, where Alice measures the input
system and communicates the outcome to Bob, who prepares the output
according to her prescription. Devising the CFT is hence a problem of {\em quantum estimation} \cite{qet}.


 In the CV setting, the CFT has been only assessed for the special
instance of pure coherent input states
 \cite{bfkjmo,hammerer}.  In
the limit case of completely unknown coherent
 amplitude, it yields
the benchmark ${\cal  F}^{cl}_{coh}=1/2$,
 which has been
extensively employed
 to validate experiments
\cite{furuscience,memorypolzik,telepolzik}.
 However, no such
threshold is known for the case of {\em squeezed} states, which are
currently drawing attention as input states of
transfer
 protocols: the high-fidelity teleportation of squeezed
states
 may enable
 cascading of teleportation, resulting in the
construction of
 non-Gaussian gates (e.g.~the cubic phase gate),
useful to achieve universal CV quantum
 computation
 \cite{fururep}. The lack of a CFT for squeezed states makes the
validation of experiments a rather controversial issue, as there is
no clear way to establish whether the achieved performances are
signatures  of a genuinely quantum information processing.

In this Letter we provide the first quantum benchmark for teleportation and
storage of single-mode squeezed states with zero displacement and
completely unknown degree of squeezing $r$ along a given direction.
This encompasses the typical experimental situation \cite{furuorig} in
which a squeezed state with known phase is generated, with the
possibility of varying the degree of squeezing by tuning the
nonlinearity of the optical parametric oscillator.  For ideally {\em
  pure} input squeezed states (and pure squeezed reconstructed
outputs \cite{notebob}) the CFT reads ${\cal  F}^{cl}_{sq(\mu =1)} \approx 81.5\%$.
We also address the search of a quantum benchmark for squeezed {\em
  thermal} states, which is necessary for a fair comparison with the
actual experiments \cite{notemixed}. We show that a moderate amount of
input mixedness only slightly modifies the classical average fidelity,
rendering our benchmark robust against thermal noise. To prove this we
provide an upper and a lower bound on the classical average fidelity,
that coincide for pure input states and are close to each other for
moderately mixed input
states. 
Thanks to these results we provide a detailed discussion of recent
experiments on teleportation and storage of squeezed (thermal) states
\cite{furuorig,braunsqz,lvovski}, which appear on the edge to beat the
upper bound, hence to pass the test of quantumness provided by our benchmark.

 We consider the transmission of single-mode non-displaced
squeezed
 thermal states $\varrho^{in}_{r,\mu}$ with purity $\mu
\equiv {\rm
 Tr}[{\varrho^{in}_{r,\mu}}^2]$ and squeezing degree $r$
along a
 fixed direction (which we can assume to be the position axis
without
 loss of generality \cite{furuorig}),
 defined by the action of the
 phase-free
squeezing operator $\hat U(r)= \exp [\frac{r}{2} (\hat
 {a}^{\dag2}
-\hat {a}^2) ]$ on a thermal state $\varrho^{th}_{\mu}=
 (1-\Lambda)~
\sum_{n=0}^{\infty} \Lambda^n |n\rangle \langle n|$
 [where $\ket{n}$
is the $n$'th Fock state, $\Lambda=(1-\mu)/(1+\mu)$
 and the mean
number of thermal photons is $\bar
 n^{th}_\mu=(1/\mu-1)/2$]:
$\varrho^{in}_{r,\mu}=\hat
 U(r)\varrho^{th}_{\mu}\hat U(r)^\dagger$.
For $\mu=1$ one recovers
 the pure squeezed vacuum state,
$\ket{\phi_r} = \hat U(r) \ket{0}$.
 A squeezed thermal state,  belonging to the family of Gaussian states \cite{ourreview},
 can be completely described by its covariance matrix (CM)  $\sig_{r,\mu}=(1/\mu){\rm
 diag}\{\exp (2r),\,\exp(-2r)\}$.
 We
aim at computing the CFT for
 teleporting states of the form
$\varrho^{in}_{r,\mu}$, with given
 $\mu$, and completely unknown
squeezing $r$. With the expression \emph{completely unknown
squeezing} we refer here to the scenario where a verifier Victor
secretly chooses a value of the squeezing parameter, and asks Alice to
transfer the corresponding state to Bob.  The verifier will
eventually assign to Alice and Bob a score equal to the fidelity
between  input and output state. The CFT  is  the maximum score that
Alice and Bob can get in this game with a classical
``measure-and-prepare'' strategy \cite{notebob}. The verifier will always
choose the  less favourable state, forcing
the two parties to adopt the strategy that maximizes the minimum
score \cite{qet}. Such a strategy has to work equally well for any
possible value of $r$ \cite{ozawa}. This scenario corresponds to the
limit case in which the input states are drawn according to  an
ideally ``flat'' prior distribution of the unknown squeezing
parameter, in analogy to the typical situation
considered for coherent states \cite{bfkjmo, hammerer}.


In order to devise the CFT for squeezed states we employ the
techniques of Ref.~\cite{giulio}, in which a method to explicitly
devise the optimal quantum measurement to estimate a completely
unknown squeezing transformation $\hat U (r)$ acting on an arbitrary
{\em pure} single-mode state $\ket {\psi}$ was provided. Such an
optimal measurement yields an estimate of the squeezing degree equal
to $r+\delta$ with probability distribution $p^{opt}_{\psi}(\delta)$
(independent on $r$ and peaked around $\delta =0$) given by
\begin{equation}\label{OptSqueez}
\begin{split}
&p_{\psi}^{opt}(\delta) = |\bra {\psi}  \eta_{\psi} (\delta)\rangle |^2,\\
&\ket {\eta_{\psi} (\delta)} = \frac 1 {\sqrt{2
\pi}}\int_{-\infty}^{+\infty} d \nu ~ e^{-i\nu \delta}
\frac{\Pi_{\nu} \ket{\psi}}{\sqrt{\bra {\psi} \Pi_{\nu} \ket{\psi}
}}~,
\end{split}
\end{equation}
where $\Pi_{\nu} = 1/(2\pi) \int_{-\infty}^{+\infty} d \lambda~
e^{i\nu \lambda}~ \hat U(\lambda)$.  Note that the optimal measurement depends on the state $\ket{\psi}$: If the optimal
measurement for the state $\ket{\psi}$ is used to estimate the
squeezing on a different state $\ket{\phi}$
\cite{nota:NotTooDifferent}
one will generally have a suboptimal
estimation, with probability distribution given by
\begin{equation}\label{OptPhiPsi}
p_{\phi, \psi} (\delta) = |\bra {\phi} \eta_{\psi} (\delta)
\rangle|^2
\end{equation}
still depending only on $\delta$, but not necessarily peaked around
$\delta =0$. The estimation in \eq{OptSqueez} is
optimal  for a whole class of different figures of merit
\cite{nota:HolevoClass}.  In particular, the fidelity between two
squeezed thermal states with CMs $\sig_{r,\mu}$ and
$\sig_{r+\delta,\mu}$ belongs to this class for any given value of
$\mu$. It  depends only on the purity $\mu$ and on the
difference $\delta$, and is given by \cite{scutaru} ${\cal
  F}_{\delta,\mu}=2 \mu ^2/[\mu ^2+\sqrt{\mu ^4+2 \cosh (2 \delta )
  \mu ^2+1}-1]$.

We discuss now the consequences of the above results for the transfer
of phase-free squeezed states. The exact CFT for {\em pure} squeezed
vacuum input states can be obtained by setting $\ket {\psi} = \ket 0$
in Eq.~(\ref{OptSqueez}), which yields the optimal probability
distribution $p^{opt}_0(\delta) = \left| \int_{-\infty}^{+\infty}
  \frac{e^{-i \nu \delta }}{(2\pi)^{5/4}} |\Gamma (1/4 + i \nu/2 )| d
  \nu \right|^2$, $\Gamma(z)$ being the Euler gamma function.
The CFT is then given by the average fidelity of
the optimal estimation strategy, and can be evaluated by numerical
integration of the expression ${\cal F}_{sq (\mu =1)}^{cl} = \int
d \delta~ p_0^{opt} (\delta)~ \mathcal F_{\delta,\mu =1}$. This leads
to the benchmark $\mathcal{
  F}^{cl}_{sq (\mu =1)} \approx 0.81517$, as anticipated.
We notice that the CFT of $81.5\%$ is sensibly higher than the
corresponding one of $50\%$ for coherent states \cite{hammerer}.  This
could be expected, as it is easier to estimate a single real parameter
than a complex amplitude. This also shows that an experimental
demonstration of genuine quantum features is much more demanding for
squeezed states than for coherent states.

Unlike the case of coherent states (whose purity is unaffected by
photon losses), to have a fair comparison with the experiments
\cite{furuorig}, it is crucial to investigate the CFT for input
squeezed states which are realistically {\em mixed} \cite{notemixed}:
this is a much harder estimation problem. In the following we derive
both an upper and a lower bound on the average CFT for input squeezed
thermal states $\varrho^{in}_{\mu,r}$ with given purity $\mu$
(associated to the experimental losses) and completely unknown degree
of squeezing $r$. With \emph{average} CFT we mean the average over $\delta$
of the fidelity between $\varrho^{in}_{\mu,r}$ and $\varrho_{\mu,
  r+\delta}$, the state prepared by Bob when estimating $r+\delta$  \cite{notebob}, maximized over all possible estimation strategies \cite{noteaverage}.
The two
bounds individuate the value of the average CFT ${\cal \bar
  F}^{cl}_{sq(\mu)}$ within a window that gets as narrower as the
purity of the state is higher, shrinking for $\mu \to 1$ onto the
exact value of the CFT for pure states.  Let us start with the upper
bound, which is easily obtained by considering the ensemble
decomposition of the thermal state as a mixture of Fock states. For
any possible measurement performed on a squeezed thermal state the
probability distribution will be $p_{th}(\delta) = (1-\Lambda)
\sum_{n=0}^{\infty} \Lambda^{n} p_n(\delta)$, where $p_n (\delta)$ is
the probability distribution for the same measurement when performed
on the squeezed Fock state $\hat U(r) \ket n$.  If we consider the
optimal estimation, with probability $\tilde p_{th}(\delta)$, we
obtain the bound: ${\cal \bar F}^{cl}_{sq(\mu)} =
\int_{-\infty}^{+\infty} \tilde p_{th} (\delta) {\cal F_{\delta,\mu}}
~ d \delta = (1-\Lambda) \sum_{n=0}^{\infty} \Lambda^n
\big(\int_{-\infty}^{+\infty} \tilde p_n (\delta) {\cal
  F_{\delta,\mu}} ~ d \delta \big) \le (1-\Lambda) \sum_{n=0}^{\infty}
\Lambda^n \big( \int_{-\infty}^{+\infty} p^{opt}_n (\delta) {\cal
  F_{\delta,\mu}} ~ d \delta\big) \equiv {\cal \bar
  F}^{up}_{sq(\mu)}$, where $p_n^{opt} (\delta)$ is the probability
distribution of the estimation that is optimal for the Fock state
$\ket n$, given by \eq{OptSqueez} with $\ket {\psi} = \ket n$.
Explicitly, $p^{opt}_n(\delta) = \left| \int_{-\infty}^{+\infty}
  \frac{e^{i \nu \delta }}{2\pi}\sqrt{I^\nu_n} d \nu \right|^2$, where
$I^\nu_n= \int_{-\infty}^{+\infty} d \lambda ~ e^{i \lambda \nu}
\langle n| \hat U(\lambda) |n\rangle$, and $\langle n| \hat U(\lambda)
|n\rangle =(\cosh \lambda)^{-n-1/2} \, _2F_1[(1-n)/2,-n/2;1;-\sinh ^2
\lambda]$, $_2F_1$ denoting the Gauss hypergeometric function.
  We stress that the upper bound is attained only for
pure input states ($\Lambda \to 0$): for $\Lambda \neq 0$ one has
${\cal \bar F}^{cl}_{sq(\mu)}<{\cal \bar F}^{up}_{sq(\mu)}$ strictly,
as the optimal measurement of \eq{OptSqueez} depends on which state
$\ket{\psi}$ is squeezed, and for a thermal ensemble there is no way
to know which squeezed Fock state is measured.

On the other hand, any measurement performed by Alice automatically
provides a lower bound on the CFT.  We devise here a suitable
estimation strategy working for squeezed thermal states.  Observing
that the squeezing transformation $\hat U(r)$ commutes with the parity
operator $\hat P=\sum_n (-1)^n \ket n \bra n$, to estimate $r$ we can
perfectly separate the Fock states with even $n$ from those with odd
$n$.  The estimation strategy works as follows: (i) perform a L\"uders
measurement of the parity, thus projecting the squeezed thermal state
onto the even/odd subspace: if the outcome is $+1$, then the state
will be proportional to $\sum_n^{even} \Lambda^n~ \hat U (r) \ket n
\bra n| \hat U^\dag (r)$, otherwise it will be proportional to
$\sum_n^{odd} \Lambda^n ~ \hat U (r) \ket n \bra n \hat U^\dag(r)$;
(ii) for parity $+1$ (even subspace), perform the measurement that is
optimal for the vacuum  $\ket 0$, otherwise perform the
one that is optimal for the one-photon state $\ket 1$.
The probability distribution obtained with this strategy is
$p_{th} (\delta) = (1-\Lambda) \left[ \sum_n^{even} \Lambda^n p_{n,0}
  (\delta) + \sum_n^{odd} \Lambda^n p_{n,1} (\delta) \right]$, with
$p_{n,0}$ and $p_{n,1}$ as in \eq{OptPhiPsi}, and yields the bound: $
{\cal \bar F}^{cl}_{sq(\mu)} \ge \int_{-\infty}^{+\infty} p_{th}
(\delta) {\cal F_{\delta,\mu}} ~ d \delta \equiv {\cal \bar
  F}^{lo}_{sq(\mu)}$. We notice that this bound also converges to the
actual CFT for $\mu \rightarrow 1$.

\begin{figure}[tb!]
\includegraphics[width=8cm]{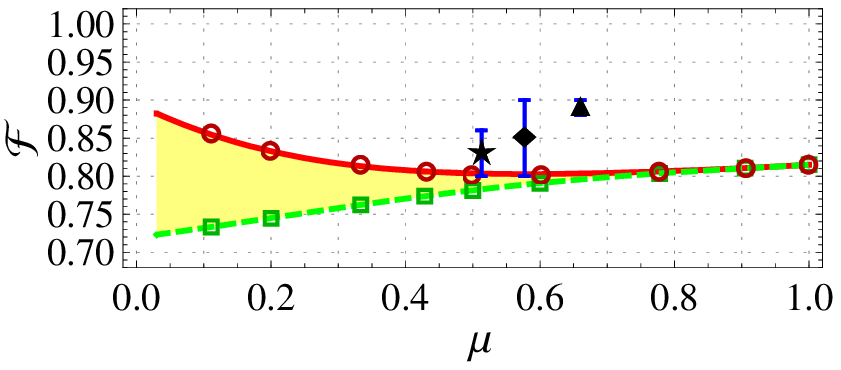}\vspace*{.1cm}
\begin{tabular}{c c c c c c c c c c c}
  \hline \hline
  $\mu$ & 1/9 & 1/5 & 1/3 & 3/7 & 1/2 & 3/5 & 7/9 & 19/21 & 1 \\
  \hline
  ${\cal \bar F}^{up}_{sq(\mu)}$ {\scriptsize{[\%]}} & 85.5 & 83.3 & 81.3 & 80.6 & 80.4 & 80.3 & 80.6 & 81.1 & 81.5 \\
  ${\cal \bar F}^{lo}_{sq(\mu)}$ {\scriptsize{[\%]}} & 73.3 & 74.5 & 76.2 & 77.4 & 78.1 & 79.1 & 80.3 & 81.1 & 81.5 \\
  \hline \hline
\end{tabular}
\caption{(color online) Plot of the average CFT  for the transfer of
single-mode  squeezed thermal states with purity $\mu$. Hollow
circles (squares), whose coordinates are reported in the table,
provide an upper (a lower) bound on the average CFT and admit a polynomial
fit by the solid (dashed) line, with equation $\frac{4 \mu
^4}{25}-\frac{11 \mu ^3}{20}+\frac{3 \mu ^2}{4}-\frac{11 \mu
   }{25}+\frac{179}{200}$
   $\big($resp. $\frac{21 \mu ^4}{200}-\frac{6 \mu ^3}{25}+
   \frac{3 \mu ^2}{25}+\frac{11 \mu }{100}+\frac{18}{25}$$\big)$.
Filled symbols correspond to the measured fidelities versus the
purity of the produced states, achieved in  recent  demonstrations
(as discussed later in the text): the diamond
``{\scriptsize{$\blacklozenge$}}'' [$\mu \approx 0.58$, $|r| \approx
5.3$ dB, ${\cal F}= (85\pm5)\%$], referring to \cite{furuorig}, and
the star ``$\boldsymbol\star$'' [$\mu \approx 0.51$,  $|r| \approx
9.1$ dB, ${\cal F}=(83\pm3)\%$], referring to \cite{braunsqz},
correspond to teleportation experiments; while the triangle
``$\blacktriangle$'' [$\mu \approx 0.66$,  $|r| \approx 5.4$ dB,
${\cal F}=(89\pm1)\%$] denotes the storage experiment of
Ref.~\cite{lvovski}.} \label{ficftemp}
\end{figure}

The upper and lower bounds on the average CFT have been numerically evaluated for
several values of input purity $\mu<1$, down to $\mu=1/9 \approx
0.11$, as plotted in Fig.~\ref{ficftemp}.  We notice that they are
very close to each other if the input states are affected by a
moderate amount of thermal noise, resulting in an error which is
smaller than $2\%$ in the experimentally relevant region of $\mu \ge
1/2$. This allows us to conclude that in this region the average CFT
slightly decreases with decreasing $\mu$ compared to the benchmark at
$\mu=1$.  The two bounds become looser in the highly mixed
regime, 
still allowing us to locate the average CFT between $70\%$ and $90\%$
for extremely thermalized input states ($\mu$ in the vicinity of $0$).

In our approach we have so far assumed the purity $\mu$ of the input
thermal state to be perfectly known (requiring an exact knowledge of
the experimental losses). However, by averaging our bounds over $\mu$
we can readily get bounds on the average CFT holding when the squeezed
thermal states prepared by Alice and Bob have equal purity randomly
distributed according to an arbitrary probability distribution
$p(\mu)$.  For example, for a flat distribution of input purity in the
range $\epsilon \le \mu \le 1, ~ \epsilon = 1/9$, we obtain the upper bound
${\cal \bar F}^{cl}_{sq} < 81.3 \%$. This and similar bounds for
$\epsilon \to 0$ can be used to discuss experiments in which both $r$
and $\mu$ are assumed to be completely unknown. The corresponding
average CFT provides a test that has to be passed by truly quantum
implementations, where $\mu$ and $r$ are not independent \cite{notemixed} (they seem related by an empirical law of the form  $\mu \propto e^{-a |r|^b}$, with suitable parameters $a,b$).

We now apply our results to the analysis of recent experiments
involving teleportation and storage of squeezed states, as shown in
Fig.~\ref{ficftemp}. Two experiments dealt with teleportation of
non-displaced squeezed thermal states of light with known phase
\cite{furuorig,braunsqz}, achieving fidelities ${\cal F}=(85\pm5)\%$
at $\mu \approx 0.58$ \cite{furuorig}, and ${\cal F} = (83\pm3)\%$ at
$\mu \approx 0.51$ (here broadband squeezing was teleported)
\cite{braunsqz,yoneprive}, respectively. The expected values for the
measured fidelities reasonably appear to pass our test, even though we
judge that the challenge of a clear-cut demonstration of quantum
teleportation of squeezed light, which likely appears within
reach, is not closed yet.  Regarding quantum memories, two
 recent experiments dealt with  storage and retrieval of
squeezed states using electromagnetically induced transparency
\cite{honda,lvovski}.  In particular, in \cite{lvovski} the fidelity
${\cal F} =(89\pm1)\%$ is reported for squeezed thermal input states
with $\mu \approx 0.66$, neatly surpassing our upper bound on the average CFT.

We anyway stress that in all discussed experiments a single squeezed
thermal state was teleported/stored.
Unambiguous demonstration of a quantum transfer would require instead
the CFT to be overcome by average experimental fidelities obtained in
experiments involving several input states with a whole (ideally, infinite) range of
values of $r$.  In any experiment in
which the input squeezing distribution is not ideally ``flat'', but
has a realistically finite width, surpassing the value of our
benchmark is a strictly {\em necessary} condition for demonstrating a
quantum
feature.
This simple observation highlights the serious difficulties of the
conventional CV teleportation protocol \cite{brakim} (see also
\cite{brareview,fururep,furuscience}), originally designed for
coherent states, when employed for input squeezed states. Let us
consider for simplicity the case of pure input states, with the
entangled resource shared by Alice and Bob being a twin-beam Gaussian
state with squeezing $s$, given by
$\ket{\Phi_s^{tb}}_{AB}=\sum^{\infty}_{n=0}[\tanh ^n(s)/\cosh (s)]
\ket{n}_A\ket{n}_B$. Then, the output of CV teleportation is still a
Gaussian state and the fidelity between input and output is ${\cal
  F}^{Q}(r,s)= \{2 e^{-2 s} [\cosh (2 r)+\cosh (2 s)]\}^{-1/2}$. We
note that, for a given resource squeezing $s$ (a measure of the shared
entanglement), ${\cal F}^{Q}(r,s)$ decreases with the input squeezing
$|r|$, vanishing in the limit $|r| \to \infty$.  Differently from the
measure-and-prepare strategy presented here, which works
equally well for any value of $r$, the performances of the
conventional quantum protocol drop exponentially with the amount of
input squeezing. For any finite $s$ there is always a critical value
$r_c$ beyond which the quantum protocol becomes less efficient than
the classical strategy, ${\cal F}^{Q}(|r|>r_c,s) < {\cal \bar
  F}^{cl}_{sq(\mu =1)}$, despite the presence of entanglement.
Moreover, using twin-beam resources with $s \le 0.74$ ($6.4$ dB), the
fidelity of the teleported states is smaller than the CFT for {\em
  any} value of the input squeezing $r$.  In short, the conventional
CV teleportation scheme \cite{brakim}, while working excellently for
coherent states (the fidelity does not depend on the complex
amplitude, and the CFT of $50\%$ is beaten iff entanglement is
shared), is not really suited for the quantum transmission of
squeezing.

The above discussion enables us to conclude that, no matter how
efficient is the setup, no experiment using conventional teleportation
can pass our test for {\em all} values $r$ of the squeezing. In view
of this, the present benchmark serves as a ``minimal test'' that has
to be passed in the finite range of values in which the experiment is
designed to successfully work.  The presented result strongly
motivates the search of new schemes that are explicitly taylored for
input squeezed states: it is very desirable to have \emph{covariant}
protocols that work equally well for any input squeezing,
and among those to determine the optimal  quantum strategy, able
to beat the classical
threshold as soon as entanglement is shared by Alice and Bob.  \\
\noindent We warmly acknowledge discussions with A.  Monras, A.
Serafini, E.  Polzik, L.  Maccone, P. Perinotti, M. F.  Sacchi, M.
Aspachs, J.  Calsamiglia, M. Piani, S. L.  Braunstein, H. Yonezawa, A.
Furusawa, and A. Lvovsky. GA was supported by Consolider-Ingenio 2010
CSD2006-0019 QOIT.  GC was supported by the Italian Ministry MIUR
through PRIN 2005.


\begin{thebibliography}{99}
\bibitem{notebob} In this Letter we always assume Alice's and Bob's
  labs to have the same ability in preparing quantum states: to
  reconstruct Alice's squeezed (thermal) states, Bob can only prepare
  squeezed (thermal) states. Allowing more freedom on Bob's side
  could further increase the CFT, leading to more demanding quantum
  benchmarks. A detailed discussion of such cases is beyond the scope of this work and will be provided elsewhere.
\bibitem{notemixed} For squeezed states, the higher the squeezing
  level is, the weaker the state becomes against losses. Such losses,
  associated to state preparation, can be experimentally estimated and
  modeled at first as input thermal noise.
\bibitem{nota:NotTooDifferent} The Dirac-normalized vectors
 $\ket
  {\eta_{\psi} (\delta)}$ provide a resolution of the
  identity
 in the subspace $\mathcal H_{\psi}$ spanned by the states $\hat U(r)
 \ket {\psi}$. The corresponding measurement can be performed on
 any state $\ket
  {\phi} \in \mathcal H_{\psi}$ (e.g. for $\ket {\psi},\ket{\phi}$ Fock states with equal parity).
\bibitem{noteaverage}
Note that the average
fidelity is smaller than the fidelity between
$\varrho^{in}_{r,\mu}$ and the average state prepared by Bob
$\varrho_{\mu,r}^{ave}= \int p(\delta) \varrho_{\mu, r+\delta}$. Therefore the experimental fidelity in a quantum implementation must exceed  our average CFT, as in Fig.~\ref{ficftemp}.
\end{thebibliography}
\end{document}